\documentclass[aps,prd,11pt,amsmath,amssymb,amsfonts,tightenlines,notitlepage,nofootinbib,superscriptaddress,floatfix]{revtex4-1}

\usepackage{epsfig,latexsym,cancel,amssymb,amsmath}
\usepackage{comment}
\usepackage{graphicx,subfigure}
\usepackage{epstopdf}
\usepackage{hyperref, graphicx, slashed, amsmath}
\usepackage[dvipsnames]{xcolor}

\usepackage{multirow}

\def\be{\begin{equation}}
\def\ee{\end{equation}}
\def\ba{\begin{eqnarray}}
\def\ea{\end{eqnarray}}
\newcommand{\beq}{\begin{equation}}
\newcommand{\eeq}{\end{equation}}
\newcommand{\bea}{\begin{eqnarray}}
\newcommand{\eea}{\end{eqnarray}}

\newcommand{\nn}{\nonumber \\}

\begin{document}

\title{Solar Reflection of Dark Matter}

\author{Haipeng An}
\affiliation{Department of Physics, Tsinghua University, Beijing 100084, China}
\affiliation{Center for High Energy Physics, Tsinghua University, Beijing 100084, China}
\affiliation{Center for High Energy Physics, Peking University, Beijing 100871, China}

\author{Haoming Nie}
\affiliation{Department of Physics, Tsinghua University, Beijing 100084, China}

\author{Maxim Pospelov}
\affiliation{William I. Fine Theoretical Physics Institute, School of Physics and Astronomy,
University of Minnesota, Minneapolis, MN 55455, USA}

\author{Josef Pradler}
\affiliation{Institute of High Energy Physics, Austrian Academy of Sciences, 1050 Vienna, Austria}

\author{Adam Ritz}
\affiliation{Department of Physics and Astronomy, University of Victoria, 
Victoria, BC V8P 5C2, Canada}

%
\begin{comment}
\author[a,b,c]{Haipeng An}
\author[a]{Haoming Nie}
\author[d]{Maxim Pospelov}
\author[e]{Josef Pradler}
\author[f]{Adam Ritz}
\affiliation[a]{Department of Physics, Tsinghua University, Beijing 100084, China}
\affiliation[b]{Center for High Energy Physics, Tsinghua University, Beijing 100084, China}
\affiliation[c]{Center for High Energy Physics, Peking University, Beijing 100871, China}
\affiliation[d]{William I. Fine Theoretical Physics Institute, School of Physics and Astronomy,
University of Minnesota, Minneapolis, MN 55455, USA}
\affiliation[e]{Institute of High Energy Physics, Austrian Academy of Sciences, 1050 Vienna, Austria}
\affiliation[f]{Department of Physics and Astronomy, University of Victoria, 
Victoria, BC V8P 5C2, Canada}
\end{comment}
%

\date{August 2021}

\begin{abstract}
The scattering of light dark matter off thermal electrons inside the Sun produces a ``fast" sub-component of the dark matter flux that may be detectable in underground experiments. We update and extend previous work by analyzing the signatures of dark matter candidates that scatter via light mediators. Using numerical simulations of the dark matter-electron interaction in the solar interior, we determine the energy spectrum of the reflected flux, and calculate the expected rates for direct detection experiments. We find that large Xenon-based experiments (such as XENON1T) provide the strongest direct limits for dark matter masses  below a few MeV, reaching a sensitivity to the effective dark matter charge of better than $10^{-9}e$. 

\end{abstract}

\maketitle

\section{Introduction}

The evidence for dark matter (DM), through its gravitational signatures on multiple astrophysical and cosmological scales, continues to stand as one of the primary motivations for physics beyond the Standard Model. As the technology of direct detection experiments improves, and experimental sensitivity to dark matter in the galactic halo is pushed down toward the threshold for elastic scattering of neutrino background fluxes ({\em i.e.} the ``neutrino floor"), there is increasing attention focused on a variety of DM models that go beyond the traditional WIMP paradigm. Coupled to the broadening scope of DM searches is the identification of ``blind spots" for direct detection, and associated experimental efforts to address them. 

Until recently, one such problematic topic in direct detection has been the realm of light WIMPs, where the momentum transfer in scattering often falls below the detection threshold for the majority of DM experiments searching for nuclear recoil. Light dark matter may exist well below the low mass end of the traditional thermal relic WIMP window \cite{Lee:1977ua}, due to the possibility of light mediators playing a role in freeze-out in the early universe \cite{Boehm:2003hm,Pospelov:2007mp,Knapen:2017xzo}. In such models, scattering of DM on electrons may provide better prospects for detection. To address the direct detection of DM in the MeV and sub-MeV mass range via its scattering on electrons, new detection technologies with ever decreasing energy thresholds (reaching down to $O({\rm eV})$ scales) are being developed \cite{Battaglieri:2017aum,SENSEI:2020dpa,DAMIC:2019dcn,CDMSLite}. At the current time, these devices are relatively small, and as a consequence operate with exposures that are many orders of magnitude smaller than the exposures achieved in the large Xenon-based experiments \cite{Aprile:2017iyp,Akerib:2016vxi,Tan:2016diz}. An alternative pathway for generating extra sensitivity is to account for sub-components of DM in full halo distribution that are considerably more energetic than the primary galactic flux, and therefore are visible to conventional underground detectors that have higher thresholds, but also large volumes and exposures \cite{Kouvaris:2015nsa,An:2017ojc,Emken:2017hnp,Bringmann:2018cvk,Ema:2018bih,Dent:2019krz,Cappiello:2019qsw,Harnik:2020ugb,Ge:2020yuf,Chen:2020gcl,Emken:2021lgc,Bell:2021xff}. These energetic sub-components have been identified after carefully considering the impact of stars, cosmic rays, and other astrophysical processes on dark matter in the halo distribution via non-gravitational scattering. 

\begin{figure}
\centering
  \includegraphics[viewport=72 555 580 760, clip=true,
  scale=0.4,width=0.65\textwidth]{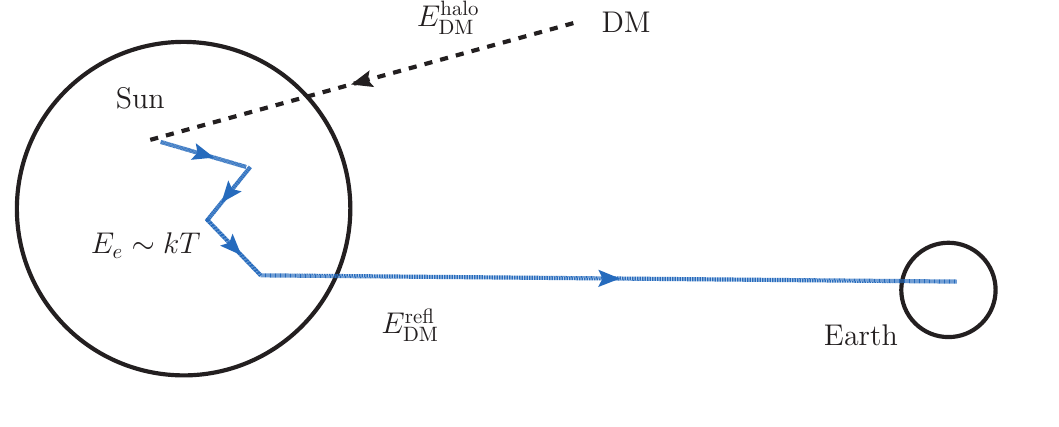}
\caption{A schematic illustration of the reflected dark matter flux generated through solar scattering.}
\label{fig:overview}
\end{figure}

In \cite{An:2017ojc}, we pointed out that there is an important direct detection sensitivity to DM in the 10 keV – 10 MeV mass range due to a population of ‘reflected DM’ generated through scattering by more energetic electrons in the Sun (or the Earth) prior to intersecting the detector. Reflective (single or multiple) scattering by solar electrons allows the kinetic energy of a sub-component of the ambient dark matter in the halo to be lifted into the keV range, making such light dark matter candidates visible above the detector energy thresholds. The scenario is illustrated in Fig.~\ref{fig:overview}, where MeV-scale DM scattering off free electrons in the Sun generates a more energetic component of the flux. While the fraction of that flux impinging on the Earth is necessarily subject to a geometric suppression factor, it was shown in \cite{An:2017ojc} that this more energetic component of the spectrum provides new sensitivity of multiple existing experiments to MeV-scale DM models.

Denoting the reduced DM-$e$ mass  $\mu_{{\rm DM},e}$, a single scatter can result in the energy of the reflected DM being raised subject to the constraint, 
\be E_{\rm
  DM}^{\rm refl} < E_{\rm DM}^{\rm refl,max} = 
\frac{4 E_e\mu_{{\rm DM},e}}{m_e+m_{\rm DM}}= \frac{4 E_e m_{{\rm DM}} m_e}{(m_e+m_{\rm DM})^2}, \label{Erec} \ee 
which is much higher than $E_{\rm DM}^{\rm halo}$ and comparable to the typical solar electron kinetic energy $E_e\sim kT_e \sim O({\rm keV})$.
Thus $E_{\rm DM}^{\rm refl}$ can be
above the detection threshold of several existing direct detection experiments. In the limit of small scattering cross sections (and a nearly transparent Sun for DM), the signal of the reflected flux scales quadratically with the scattering cross section $\propto \sigma^2$, as two scattering events, inside the Sun and inside the detector, are involved. 

The reflection mechanism is most effective for dark matter with a mass of ${\cal O}({\rm MeV})$, {\em e.g.} comparable to the electron mass. It is important to keep in mind that DM models with mass in the MeV or sub-MeV mass range are subject to a number of cosmological constraints from BBN and the CMB if they ever achieved thermal equilibrium with Standard Model degrees of freedom in the universe. In this paper, we extend the analysis of \cite{An:2017ojc} to include benchmark sub-thermal MeV-scale models of dark matter interacting via a light vector mediator and with extremely small effective coupling to electrons. Very small couplings may actually be responsible for the DM abundance via the so-called ``freeze-in" mechanism.  Calculation of the reflected flux in this case requires a somewhat different analysis of scattering inside the Sun, due to the longer-range interaction, which allows us to model DM scattering in using the model of a millicharge particle propagating through the solar electron-ion plasma. We assess the sensitivity of the Xenon1T experiment to the reflected DM component in such scenarios, and compare these results with experiments that have achieved the lowest energy thresholds for detection ({\em  e.g.} SENSEI \cite{SENSEI:2020dpa}). 

Calculation of the experimental sensitivity to Reflected DM involves two primary steps that we will describe in this paper; (i) computation of the spectrum of reflected DM from the Sun, for which we use a simulation described in Sections II, III and IV to describe scattering mediated by both heavy and light mediators; and (ii) the computation of the experimental direct detection sensitivity to DM-electron scattering, which is somewhat novel as the DM can be semi-relativistic, and our approach is described in Section V. We present our final sensitivity results for DM models with both heavy and light mediators in Section 5, and finish with some concluding remarks in Section VI.

\section{Overview of DM scattering inside the Sun}
\label{sec:strategy}

The thermal plasma inside the Sun is composed of electrons and ions, with temperatures from about 1 eV at the surface to about 1 keV at the center, and for the purpose of this study it will be safe to assume that it is known exactly. 
To fix our notation, we write the general scattering rate for DM inside the plasma as follows
\bea\label{eq:1}
\Gamma_\chi = \frac{1}{2k_1^0} \int\frac{d^3k_2}{ 2 k_2^0 (2\pi)^3} \int\frac{d^3p_1}{ 2 p_1^0 (2\pi)^3} f(p_1)\int\frac{d^3p_2}{ 2 p_2^0 (2\pi)^3} (2\pi)^4 \delta^4( k_1 + p_1 - k_2  - p_2) \sum_{\rm spin}|{\cal M}|^2 \ ,
\eea
where $k_1$ and $k_2$ are the four-momenta of the incoming and outgoing DM particle, $p_1$ and $p_2$ are the four-momenta of the incoming and outgoing electron or ion, $f$ is the momentum distribution of the electrons or the ions, and ${\cal M}$ is the scattering matrix element.
The sum is over the final state spins, and the average is performed over the initial spins. 
We will assume that the electron and ion number densities at each value of the solar radius $r$ are distributed classically according to Maxwell-Boltzmann statistics:
\bea
n_{e,{\rm ion}}(r) =\int  \frac{d^3p}{(2\pi)^3} f_{e, {\rm ion}}(p,r) \ ,
\eea
where
\bea
f_{e,{\rm ion}}(p,r) = n_{e,{\rm ion}}(r) \left(\frac{2\pi}{m_{e,{\rm ion}} T(r)}\right)^{3/2} e^{-p^2/2m_{e,{\rm ion}} T(r)} \ .
\eea
Here $n_{e,{\rm ion}}(r)$ is the 
radius-dependent number density that is given by the standard solar model. We will suppress the explicit dependence on the radius below.

Our goal is to obtain the differential scattering rate $d\Gamma_\chi / dk_2 d\cos\theta$ in the plasma frame as a function of the incoming DM momentum $k_1$, where $\theta$ is the scattering angle,
at each value of the solar radius. With this information, we can simulate the propagation of DM particles inside the Sun with the following strategy : we first assume that the plasma distribution inside the Sun is isotropic. We then discretize the radial direction into approximately ${ N}_{\rm shell}=2000$ shells, and take the temperature and density in each shell as constant. We further assume that the shells are thin enough that the probability for DM to have one collision in each shell is much smaller than unity; this assumption is justified a posteriori for the cross sections of interest.
The initial velocity of DM is a random variable that samples a truncated Gaussian distribution. 
The width of the Gaussian $\exp(-v^2/(2v_0^2)$ is given approximately by 
$v_0\sim 10^{-3}\times c$, 
and the high-velocity truncation occurs at the nominal escape velocity of $\sim 2\times 10^{-3}c$. The reflected spectrum is not sensitive to this assumption. 
The impact parameter is chosen randomly from 0 to $4R_{\odot}$ to incorporate gravitational focusing.

Outside the Sun, the effect of gravitation is accounted for through angular momentum and total energy conservation laws. Once inside the Sun, within each shell, we assume the DM particle trajectory is a straight line. Gravity is accounted for by calculating the change of velocity due to the force exerted by the inner shells. Within each shell, with the temperature and electron and ion densities taken from the solar model, we  calculate the total and differential scattering rates. The total probability determines if scattering occurs in that shell.
If scattering does not occur, the DM particle propagates to the next shell on a straight trajectory. Gravitational bending of the trajectory is accounted for at the boundary with the next shell,
where the DM velocity $\vec{v}_\chi$ is changed 
according to the gravitational influence of the inner shells. Once scattering occurs in a shell, according to the total probability, the differential scattering rate is used to randomly generate the the new angle and momentum. This new after-scattering velocity is used to propagate DM particle to the boundary of the next shell, and so forth. The simulation continues until the DM particle leaves the Sun with a positive total kinetic energy. If the total kinetic energy is negative, it will reenter the Sun and repeat the above procedure. Finally, after shooting a sufficient number of DM particles, we can statistically determine the energy distribution of the reflected DM flux. 

The final step in the simulation is to form the normalized distribution $F_{A_\rho}$ of the DM flux traversing the ``impact disc" of area $A_\rho = \pi(4R_\odot)^2$, $\int dE_\chi F_{A_\rho} = 1$. The reflected flux observed at Earth's location is then determined \cite{An:2017ojc} as follows, 
\begin{equation}
    \frac{d\Phi_{\rm reflected}}{dE_\chi} = \Phi_{\rm halo} 
    \times \frac{F_{A_\rho}A_\rho}{4\pi({\rm A.U.})^2},
\end{equation}
where the $A_\rho/ (4\pi({\rm A.U.})^2)$ ratio contains the main geometric suppression of the reflected flux at the Earth. In addition, the kinetic energy of the reflected particle at the Earth's location is reduced according to the change in the potential energy.

\begin{figure}
\centering
\includegraphics[width=0.5\textwidth]{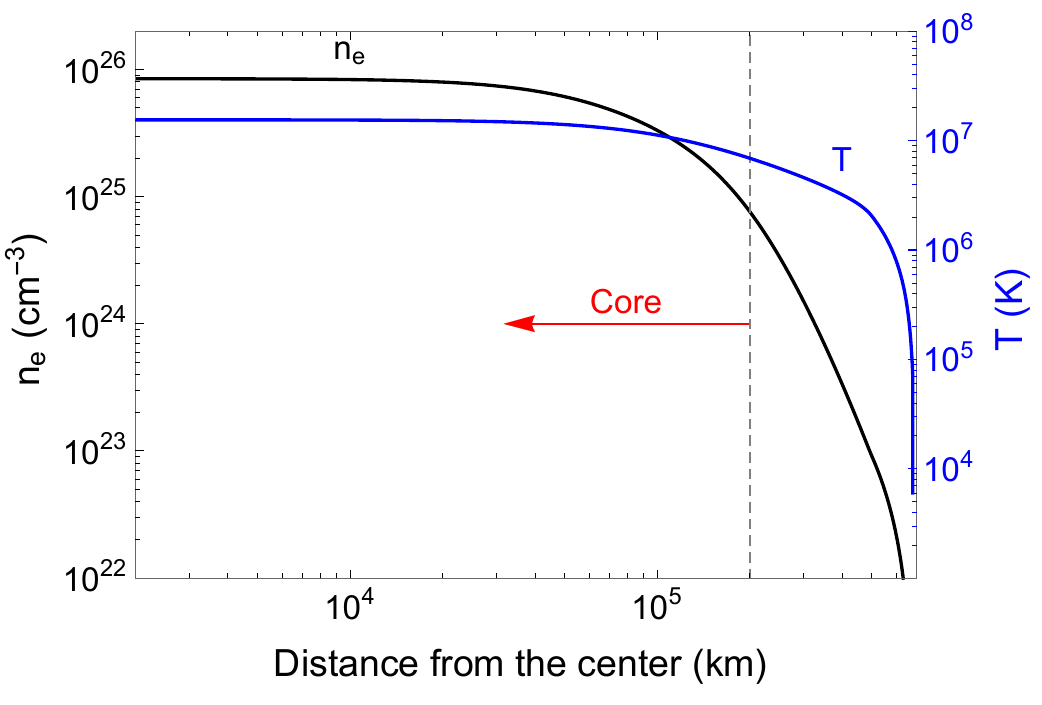}%
\includegraphics[width=0.5\textwidth]{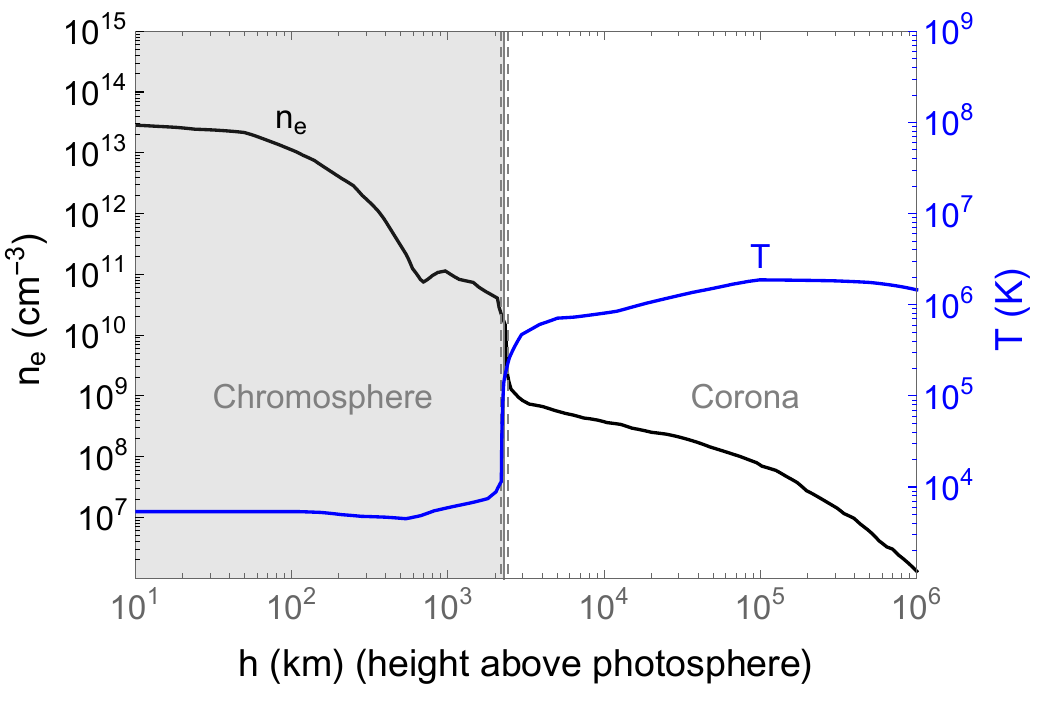}
\caption{\textit{Left panel:} temperature and electron density distributions inside the Sun taken from Ref.~\cite{Bahcall:2004pz}. \textit{Right panel:} temperature and electron density distributions in the Sun's corona reproduced from Ref.~\cite{2008GeofI..47..197D}.}
\label{fig:solarInfo}
\end{figure}

\section{Reflected DM flux: scattering via contact interactions}

In this section we describe in detail how to simulate the reflected energy spectrum in the case of contact ($s$-wave) scattering. This treatment follows previous work \cite{An:2017ojc,Emken:2021lgc}. Taking DM-electron scattering as the example, we can write the momentum-independent scattering matrix element in terms of the total scattering cross section $\sigma_{\rm tot}$, %
\bea
\overline{|{\cal M}|^2} = 16\pi (m_\chi + m_e)^2 \sigma_{\rm tot} \ .
\eea
Defining the momentum transfer as $q = k_2 - k_1$, and replacing $d^3k_2$ by $d^3q$ we obtain,
\bea
\Gamma_{\chi-e} = \frac{\pi(m_\chi+m_e)^2 \pi \sigma_{\rm tot}}{m_e^2 m_\chi^2} \int\frac{d^3p_1}{(2\pi)^3} f_e(p_1)\frac{d^3q}{(2\pi)^3} \delta\left( q^2 \left( \frac{1}{2m_\chi} + \frac{1}{2m_e} \right) + \frac{{\bf k}_1\cdot {\bf q}}{m_\chi} - \frac{{\bf p}_1 \cdot {\bf q}}{m_e}\right) \ .
\eea
Integrating out the angular part of $\vec p_1$ first, we obtain
\bea
\Gamma_\chi =  \frac{(m_e+m_\chi)^2 \pi \sigma_{\rm tot}}{m_e^2 m_\chi^2}\int \frac{d^3q}{(2\pi)^3} \int_{p_{1{\min}}} \frac{p_1^2 dp_1}{(2\pi)^2} f_e(p_1) \frac{m_e}{p_1 q} \ ,
\eea
where 
\bea
p_{1,\min} = \left| q\left(\frac{m_e}{2m_\chi} + \frac{1}{2}\right) + \frac{m_e}{m_\chi} \frac{{\bf k}_1\cdot {\bf q}}{q} \right| \ .
\eea
Given the Maxwell-Boltzmann form of the distribution $f_e$, the integral over $p_1$ can be carried out analytically so that the scattering rate takes the form:
\bea\label{eq:9}
\Gamma_\chi = \frac{(m_e+m_\chi)^2 \pi \sigma_{\rm tot} n_e m_e }{ m_e^2 m_\chi^2} \left( \frac{2\pi}{m_e T} \right)^{1/2} \int\frac{d^3q}{(2\pi)^3} \frac{1}{q} \exp\left[ - \frac{1}{2m_e T} \left( \frac{m_e}{2m_\chi} (2k_1 \cos\theta_q + q) +\frac{q}{2} \right)^2 \right]\ . \nn
\eea
It follows that the differential rate is given by
\bea\label{eq:10}
\frac{d\Gamma_\chi}{dq d\cos\theta_q} \sim {\cal N}q \exp\left[ - \frac{1}{2m_e T} \left( \frac{m_e}{2m_\chi} (2k_1 \cos\theta_q + q) +\frac{q}{2} \right)^2 \right]  \ ,
\eea
where ${\cal N}$ is a factor that does not dependent on the kinematic  variables of the process. 
The total scattering rate can then be calculated directly from Eq.~(\ref{eq:1}),
\bea
\Gamma_{\chi} = \sigma_{\rm tot} n_e \sqrt{\frac{T}{m_e}}  \times \sqrt{\frac{1}{2\pi}} \left[ 2e^{-x_\chi^2/2} + (2\pi)^{1/2} (1+x_\chi^2) {\rm erf}\left( \frac{x_\chi}{\sqrt{2}} \right) \right] \ ,
\eea
where $x_\chi = v_\chi (m_e/T)^{1/2}$. 
The probability for a DM particle to scatter off an electron in an infinitesimal distance $\Delta l$ can then be written as
\bea\label{eq:8p}
P = \Gamma_{\chi} \times \frac{\Delta l}{v_\chi} = \sigma_{\rm tot} n_e \Delta l \left[ \sqrt{\frac{\pi}{2}} \frac{1}{x_\chi} e^{-x_\chi^2/2} + \frac{1+x_\chi^2}{x_\chi^2} {\rm erf}\left(\frac{x_\chi}{\sqrt{2}}\right) \right] \ .
\eea

\begin{figure}
\centering
\includegraphics[height=3in]{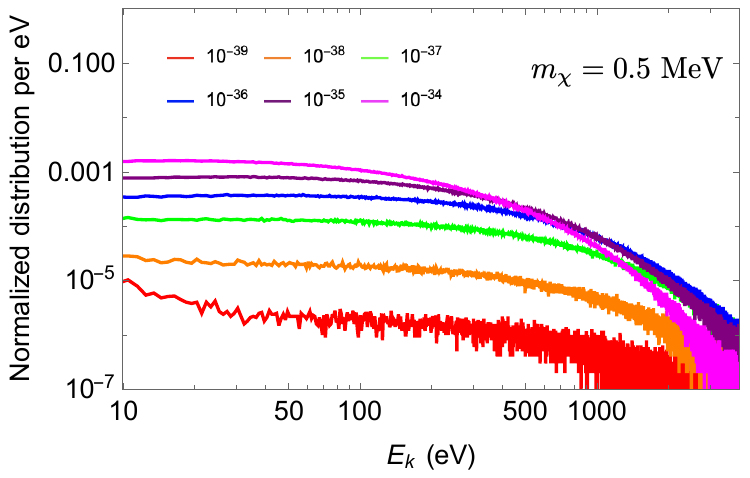}
\caption{Reflected DM energy spectrum for $m_\chi = 0.5$ MeV.
The distributions result from cross sections ranging from $\sigma_{\rm tot} = 10^{-39}$ -- $ 10^{-34}$cm$^2$. }
\label{fig:1}
\end{figure}

\begin{figure}
\centering
\includegraphics[height=3in]{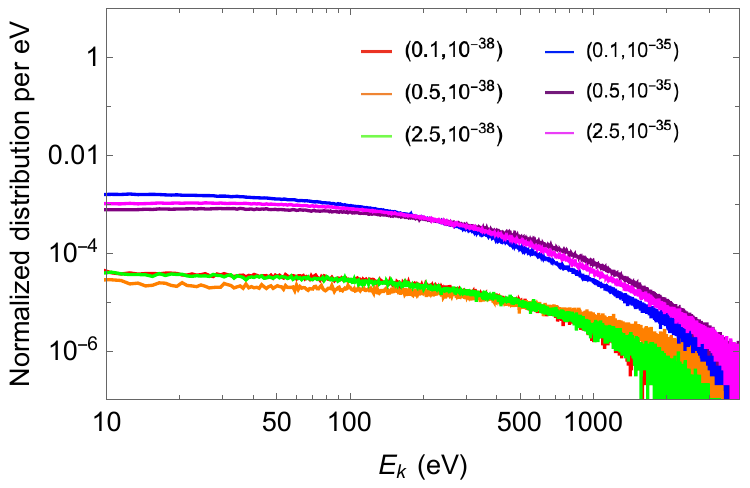}
\caption{Reflected DM energy spectrum when both DM mass and cross section are varied. The first number in paranthesis corresponds to $m_\chi$ in MeV, and the second number to 
$\sigma_{\rm tot}$ in cm$^2$. 
}
\label{fig:2p}
\end{figure}

With the differential scattering rate (\ref{eq:10}) and total scattering probability (\ref{eq:8p}) in hand, the reflected flux can be obtained following the strategy presented in Sec.~\ref{sec:strategy}. 
In Fig.~\ref{fig:1} we show the resulting normalized flux for $m_\chi = 0.5$ MeV. The DM-electron cross section is varied  from $10^{-39}$ to $10^{-34}$ cm$^2$. As the cross section increases, the DM particles experience scattering with the electrons in the outer, more dilute layers of the Sun. Consequently, the spectrum becomes softer. In Fig.~\ref{fig:2p}, we compare the reflected spectra for different values of $m_\chi$. In the non-relativistic limit, the energy transfer-efficiency reaches its maximum when the masses of the elastically scattering particles are equal. As seen in Fig.~\ref{fig:2p}, the spectra for $m_\chi = 0.5$~MeV are harder than for $m_\chi = 0.1$ and $2.5$~MeV. This conclusion applies to both large and small values of the cross section. 

\subsection{The single scattering limit and the validation of the simulation} 
\label{sec:onescatter}

It is clear that when the value of the cross section is sufficiently small, the probability of multiple scattering diminishes, and the single scattering limit adequately describes the reflected spectrum. In this subsection, we compare our simulations with a semi-analytic single-scattering formula in the appropriate limit. 

We can estimate the relevant cross section for the single-scattering approximation by considering the scattering probability in the solar 
core. The relevant mean free path is $l_{\rm fp} = [n_e \langle \sigma_{\rm tot} v_r\rangle]^{-1} {\bar v_\chi}$ where ${\bar v_r} \sim {\bar v_e}$ and ${\bar v_\chi}$ are respectively the characteristic relative velocity (approximately the electron velocity) and the DM velocity in the core. We can then estimate the scattering probability as
\bea
P_s \sim R_{\rm traj} / l_{\rm fp} \sim \rho_{\rm core}/m_p \times  R_{\rm traj} \times \sigma_{\rm tot} \times \frac{\bar v_e}{\bar v_\chi} \sim  \frac{\sigma_{\rm tot}}{10^{-38}~{\rm cm}^2} \ , 
\eea
where $\rho_{\rm core}$ is approximately $150$ gram$/$cm$^3$. Inside the core, the typical ratio of electron to DM velocities ${\bar v_e}/{\bar v_\chi} \sim 10$. $R_{\rm traj}$ refers to the typical distance that a DM particle travels inside the core of the Sun, where the core size is about one quarter of the solar radius.

\begin{figure}
\centering
\includegraphics[height=3in]{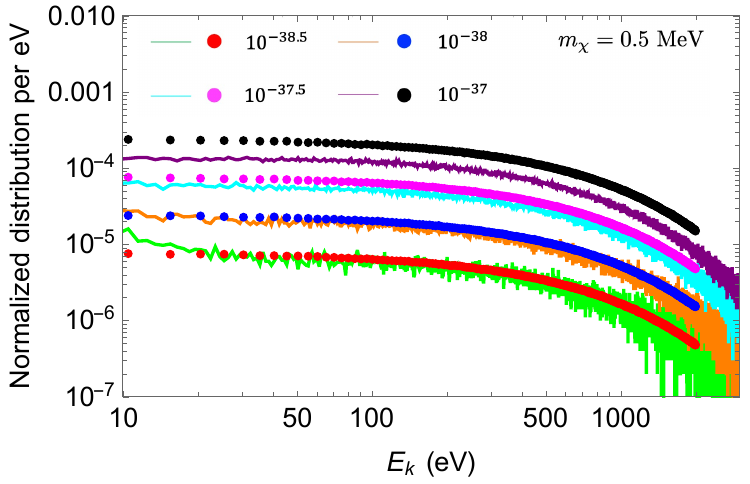}
\caption{Comparison of the single-scattering approximation (heavy dots) with numerical simulations (scatter points) for $m_\chi =0.5$ MeV and different values of the cross section in units of cm$^2$. }\label{fig:2}
\end{figure}

Therefore, in the region where $\sigma_{\rm tot} \ll 10^{-38}$ cm$^2$, the probability of scattering for each DM particle traversing the Sun is much smaller than one, and the reflected flux can be estimated by considering just single scattering events. For such events, the DM trajectory prior scattering is determined by  gravity alone. As a result, the differential scattering probability simplifies to
\bea\label{eq:310}
\left.\frac{dP}{d E_2}\right|_{\rm trajectory} = \int_{\rm trajectory} \frac{d l}{v_\chi} \frac{d\Gamma_\chi}{d E_2} (k_1(r(l)), T(r(l)), n_e(r(l)))\ ,
\eea
where $v_\chi$ is the speed of the DM particle at the moment of scattering. The trajectory of the integral is solely determined by the initial velocity and impact parameter;  $E_2$ is the kinetic energy of the outgoing DM particle. In the region of interest, we have $k_2 \gg k_1$, and as a result $q\approx k_2$. In this limit, we may integrate  $d\cos\theta$ in Eq.~(\ref{eq:10}) and obtain 
\bea\label{eq:311}
\frac{d\Gamma}{d E_2} = \frac{\sigma_{\rm tot} n_e }{(2\pi m_e T)^{1/2}} \frac{(m_e+m_\chi)^2}{m_e m_\chi} \exp\left[- \frac{E_2}{T} \frac{(m_e + m_\chi)^2}{4 m_e m_\chi} \right] \ .
\eea
One observes that in this limit, the differential rate is insensitive to $k_1$. The normalized kinetic energy distribution of the reflected flux can then be written as
\bea
\frac{2}{R_{\rm max}^2 \bar v_0} \int dR\, R  \int d v_0~v_0 f_\chi(v_0)  \left.\frac{dP}{d E_2}\right|_{\rm trajectory}  \ .
\eea
In this expression, $\bar v_0$ is the average of $v_0$, $R_{\rm max}$ is the maximum value of the impact parameter, which we choose to be $4R_{\rm Sun}$ so as to include the gravitationally focused component, and $f_\chi$ is the initial DM velocity distribution. If the trajectory does not intersect with the Sun, the total kinetic energy does not change. In Fig.~\ref{fig:2} we show the normalized distributions for $m_D = 0.5$~MeV with $\sigma_{\rm tot}$ varied from $10^{-38.5}$~cm$^2$ to $10^{-37}$~cm$^2$. The colored curves show the result using the  simulation described in Sec.~\ref{sec:strategy}. The big dots show the result obtained using the single scattering calculation. We observe that for $\sigma_{\rm tot} \ll 10^{-37}$ cm$^2$, the single scattering results agree  with the simulation. For larger values of $\sigma_{\rm tot}$ one leaves the single-scattering regime and the analytical approximation begins to over-estimate the reflected flux. 

\subsection{Including the effect of ions}

The DM particles may also interact with ions, and the strength of the corresponding cross section
relative to the cross section on electrons is model-dependent. While it is fixed for scattering via a dark photon type mediator, it will be different for other mediators, in general. 
Since we consider $m_\chi \ll m_{\rm nucleus}$, collisions with ions will only change the direction of DM. If the collision happens when the DM particle goes inward to core, the ions may shield  DM  from going into the core. However, if the collision happens when the DM particle is on an outward trajectory without having been accelerated non-gravitationally, the collision may turn DM back into the core and enlarge its probability for up-scattering. Without a detailed simulation it is therefore hard to discern whether ions enhance or diminish the reflected flux. 

\begin{figure}
\centering
\includegraphics[height=3in]{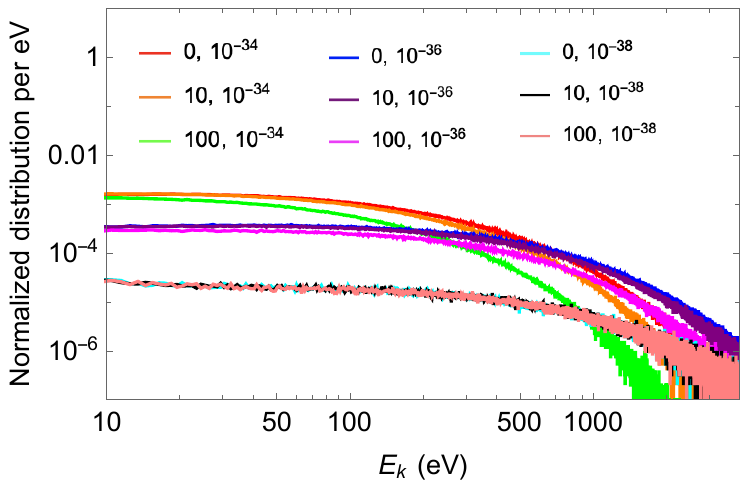}
\caption{Reflected spectrum upon inclusion of ions for $m_\chi$=0.5\,MeV. The first number in the legend is $\sigma_{\rm tot}^{\chi-{\rm ion}}/\sigma_{\rm tot}^{\chi-{e}}$ and the second number is $\sigma_{\rm tot}^{\chi-e}$ in units of cm$^2$. }\label{fig:ion}
\end{figure}

The average velocity of ions inside the core of the Sun is about $10^{-3}$ times of the speed of light, and about one order of magnitude smaller than the typical velocity of DM inside the core. Therefore, the scattering rate is determined by the velocity of DM. As a result, relative to $\chi-$e scattering, $\chi-$ion scattering suffers from an approximate suppression factor of $v_{\chi} / v_e \sim 0.1$. Therefore, when the total scattering probability on electrons is well below unity (single-scattering limit), ions cannot change the reflected flux. In Fig.~\ref{fig:ion}, we show the energy distributions of the reflected flux for different choices of $\sigma_{\rm tot}^{\chi-e}$ and $\sigma_{\rm tot}^{\chi-{\rm ion}}$. For each curve, the first number in the legend shows the ratio $\sigma_{\rm tot}^{\chi-p}/\sigma_{\rm tot}^{\chi-{e}}$ and the second number shows the value of $\sigma_{\rm tot}^{\chi-{e}}$; 
$m_\chi$ is chosen to be $0.5$~MeV. We observe that when $\sigma_{\rm tot}^{\chi-e}$ is below roughly $10^{-38}$~cm$^2$, the ions do not have any noticeable impact. On increasing $\sigma_{\rm tot}^{\chi-e}$, ions suppress the high-energy tail of the reflected spectrum. Therefore, we conclude that the net effect of ions is to shield  DM from the core.

\section{Reflected DM flux: scattering via light mediators}

For reasons that we will discuss at length, DM scattering via a very light mediator requires a more involved calculation of the reflected flux compared to the case of contact interactions. 
Here we assume that the dark mediator has a negligible mass compared to the momentum 
transfers considered, which in 
practice amounts to a sub-eV range.\footnote{In some models it might be important that we keep the mediator mass above a specific low threshold; for example, the propagation of millicharged particles is affected by magnetic fields, solar winds etc \cite{Dunsky:2018mqs}. A small dark photon mass above $10^{-13}$eV, on the other hand, would suppress the effect of the magnetic field beyond distances of 1000 km regardless of the effective charge.}
There are  two principal options to realize an almost massless dark mediator: one is through kinetic mixing with the photon and the other is through a new force, e.g.~a vector boson coupled to the $U(1)_{e-\mu}$ or $U(1)_{e-\tau}$ currents with a tiny coupling. In this section, we will discuss these two cases separately. 

\subsection{Light mediator through kinetic mixing}

The Lagrangian in this case takes the form, 
\bea
{\cal L} = -\frac{1}{4} F_{\mu\nu} F^{\mu\nu} - \frac{\kappa}{2} F_{\mu\nu} V^{\mu\nu}  -\frac{1}{4} V_{\mu\nu} V^{\mu\nu} - \frac{1}{2} m_V^2 V_\mu V^\mu + V_\mu J_\chi^\mu +  A_\mu J^\mu \ ,
\eea
where $J_\chi^\mu$ is the dark sector current, and the corresponding charges are subsumed into 
the definitions of currents. Since in this paper the
symbol $\varepsilon$ is reserved for the electric permittivity, we prefer to denote the dark photon and the kinetic mixing parameter as $(V,\kappa)$ as opposed to the more common notation of $(A',\varepsilon)$. We will refer to the charge of DM under the $V_\mu$ mediator as~$e_D$. 

In a general Feynman gauge, the propagator of the photon can be written as
\bea
\Pi^{\mu\nu} = \frac{1}{\varepsilon_L({q^0}^2 - q^2)} \epsilon_L^\mu \epsilon_L^\nu + \frac{1}{\varepsilon_T{q^0}^2 - q^2} \sum_r \epsilon_{T r}^{\mu} \epsilon_{T r}^{\nu} + \frac{q^\mu q^\nu}{\xi({q^0}^2 - q^2)^2} \ .
\eea
In this expression, $\varepsilon_{L(T)}$ denotes the longitudinal (transverse) electromagnetic permittivity, $q=|{\bf q}|$,
and the summation is over transverse indices. 
The last term, the gauge-dependent part of the propagator, does not give a contribution to the scattering matrix element. In predominantly Coulomb-type scattering, it is the 0-component of $J^\mu$ that contributes the most.  For non-relativistic (NR) scattering, the matrix element can be written as
\bea
{\cal M} &=& \langle J_\chi^0 \rangle \frac{1}{q^2 - m_V^2} \kappa q^2 \frac{{\epsilon_L^0}^2}{\varepsilon_L({q^0}^2 - q^2)} \langle J^0\rangle \nn
&=& \frac{\kappa q^2}{ \varepsilon_L q^2 (q^2 - m_V^2) } \langle J_\chi^0 \rangle \langle J^0\rangle \ .
\eea
Given that in NR scattering the energy transfer is subdominant to the momentum transfer, $q^0 \ll |\vec q|$, we have
\bea\label{eq:Mlight}
{\cal M} \approx \frac{\kappa \langle J_\chi^0 \rangle \langle J^0\rangle}{\varepsilon_L (q^2 + m_V^2)} \ .
\eea
It is therefore important to calculate the longitudinal permittivity~$\varepsilon_L$. We will do so using linear response theory, the details of which are presented in Appendix~A. 

To present the results, let us introduce a dimensionless quantity $A$, 
\begin{equation}
    \label{eq:Ae_main}
A = \frac{q^0}{q} \left(\frac{m_e}{T}\right)^{1/2},
\end{equation}
in terms of which, the longitudinal permittivity is given by  
\bea\label{eq:epsilonL_main}
\varepsilon_L &=& 1 + \frac{e^2 n_e}{T q^2} \left[1 - 2A \int \frac{x dx}{(2\pi)^{1/2}} \tanh^{-1} (x/A) e^{-x^2/2} \right] \ . \nn
&\equiv& 1 + \frac{e^2 n_e}{T q^2} F_1(A)
\eea
where function $F_1$ is defined as
\bea\label{eq:F1}
F_1(A) = 1 - \left(\frac{\pi}{2}\right)^{1/2} A e^{-A^2/2} \left[{\rm erfi}(A/\sqrt{2}) - i\right] \ .
\eea
In the system of units used in this paper $e^2 = 4 \pi \alpha$. 
Not surprisingly, the main dimensionful parameter controlling the permittivity is the Debye length 
$\ell_{\rm Debye}$ defined here as $\mu_{\rm Debye}^2 = \ell_{\rm Debye}^{-2} = 4\pi\alpha n_e/T$.

The corresponding result for the transverse permittivity is given by 
\bea
\varepsilon_T = 1 - \frac{e^2 n_e}{T q^2} \frac{1}{A} \left(\frac{\pi}{2}\right)^{1/2} e^{-A^2/2} \left[{\rm erfi}\left(\frac{A}{\sqrt{2}}\right) - i\right]= 
1 + \frac{e^2 n_e}{T q^2} \frac{1}{A^2}(F_1(A)-1) \ .
\eea
As an explicit check of this result, one can take the limit where the photon is nearly on-shell ($A\gg 1$) to confirm that both permittivities have a common limit, $\varepsilon_{L(T)}\to 1 - \omega^2_p/{q^0}^2$, with the usual non-relativistic expression for plasma frequency, 
$\omega_p^2 = 4\pi \alpha m_e^{-1}n_e$.

The above result can easily be generalized to include ions, leading to the general form, 
\bea
\varepsilon_L = 1 + \frac{\mu_{\rm Debye}^2}{q^2} \left[ F_1(A_e) + \sum_i\frac{Z_i^2n_i}{n_e} F_1(A_i) \right] \ ,
\eea
where 
\bea
A_i  = \frac{q^0}{q} \left( \frac{m_i}{T} \right)^{1/2} \  ,
\eea
and $Z_i$, $m_i$ and $n_i$ are the charges, masses and  number densities of the ions. As a result, the effect of plasma screening differs if the target in the collision is different, which is a signature of model dependence. Furthermore, charged particles with different masses also have different screening effects.

\subsubsection{Screening effects in $\chi$-e and $\chi$-ion collisions}

In the case of a $\chi$-e collision, and for the hard collisions that we are interested in, 
\bea
q^0 \sim T, \;\; q \sim (T m_e)^{1/2}. 
\eea
As a result, $A_e$ is of order one. For the contribution from ions, since $m_i\sim$ GeV, we have $A_i \gg 1$, and as discussed in in the previous subsection, 
\bea
F_1(A_i) \rightarrow - \frac{1}{A_i^2} \ll 1 \ ,
\eea
Therefore, for $\chi-e$ collisions, the contribution from ions to $\varepsilon_L$ is suppressed by a factor of ${\cal O}(10^{3})$ relative to the contribution from electrons, and is therefore  negligible.

In the case of $\chi-$ion collisions we have $q^0/q \sim v_i$, where $v_i$ is the typical speed of the ions. Therefore, for the contribution from the electrons, we have $A_e \sim v_i/v_e \ll 1$. As a result, and as discussed above, $F_1(A_e) \approx 1$, and the screening contribution from electrons can be described by regular Debye screening. For the contributions from the ions, it is easy to see that $A_i$ can be of order unity. Therefore we need to rely on the full result for $F_1$ as shown in Eq.~(\ref{eq:F1}).

\subsection{Scattering rates} 

\subsubsection{$\chi-$e scattering} 

From Eqs.~(\ref{eq:1}) and (\ref{eq:Mlight})  the scattering rate can written as
\bea\label{eq:Gammachi0}
\Gamma_{\chi-e} &=& \int \frac{d^3k_2}{(2\pi)^3} \int\frac{d^3p_1}{(2\pi)^3} f_e(p_1) \frac{\kappa^2 e_D^2 e^2 q^4 }{|q^2 + \mu_{\rm Debye}^2 F(q^0/q)|^2 (q^2 + m_V^2)^2} \nn
&&\times 2\pi \delta\left(\frac{(k_1^2 - k_2^2)}{2m_\chi} + \frac{p_1^2 - (\mathbf{k}_1+\mathbf{p}_1 - \mathbf{k}_2)^2}{2m_e}\right) \ ,
\eea
where $\mu_{\rm Debye}^2 = e^2 n_e/T$ is the Debye screening parameter, and 
\bea
F(q^0/q) = F_1(A_e) + b_p F_1(A_p) + 4b_{\rm He} F_1(A_{\rm He}) \approx F_1(A_e) \ ,
\eea
where $b_{p,{\rm He}} = n_{p,{\rm He}}/n_e$, and 
$F_1(A)$ is defined as $\Delta\varepsilon_L/\Delta\varepsilon_L^{(0)}$; its real and imaginary parts are shown in Fig.~(\ref{fig:debye}). 

Following the derivation in the contact interaction case, the $e-\chi$ scattering rate can be written as
\bea
\Gamma_{\chi-e} = \frac{\kappa^2 e^2 e_D^2 n_e}{(2\pi)^{3/2}} \left(\frac{m_e}{T}\right)^{1/2}\!\!\int dq q d\cos\theta_{qk_1} \frac{1}{|q^2 + \mu_{\rm Debye}^2 F_1(A_e)|^2} \frac{q^4}{|q^2 + m_V^2|^2} \exp\left({-\frac{p_{1\rm min}^2}{2 m_e T}}\right), \qquad
\eea
where
\bea
p_{1\rm min} &=& \left|q\left(\frac{1}{2} + \frac{1}{2} \frac{m_e}{m_\chi}\right) + k_1\left(\frac{m_e}{m_\chi}\right) \cos\theta_{qk_1}\right|, \nn
A_e &=& \frac{m_e}{m_\chi} \left( \frac{k_1}{\sqrt{m_e T}} \cos\theta_{qk_1} + \frac{1}{2} \frac{q}{\sqrt{m_e T}} \right).
\eea
The rate $\Gamma_{\chi-e}$ can be simplified as follows, 
\bea\label{eq:GammaMassless}
\Gamma_{\chi-e} = \frac{n_e \kappa ^2 e^2  e_D^2}{(2\pi T)^{3/2} m_e^{1/2}} \int dx_q dc_q \exp\left[ - \frac{1}{2}\left(\frac{1}{2}x_q + A_e\right)^2 \right] \frac{1}{|x_q^2 + \frac{\mu_{\rm Debye}^2}{m_e T} F_1(A_e)|^2} \frac{x_q^5}{(x_q^2 + \frac{m_V^2}{m_e T})^2}, 
\eea
where 
\bea
A_e = v_1\left( \frac{m_e}{T} \right)^{1/2} c_q + \frac{1}{2} \frac{m_e}{m_\chi} x_q \ ,
\eea
with $c_q = \cos\theta_{qk_1}$.

\subsubsection{$\chi-$ion scattering} 

In the case of the $\chi-$ion scattering, we have
\bea\label{eq:Gammachi_ion}
\Gamma_{\chi-{\rm ion}} &=& \int \frac{d^3k_2}{(2\pi)^3} \int\frac{d^3p_1}{(2\pi)^3} f_{\rm ion}(p_1) \frac{\kappa^2 Z^2e_D^2 e^2 q^4 }{|q^2 + \mu_{\rm Debye}^2 F(q^0/q)|^2 (q^2 + m_V^2)^2} \nn
&&\times 2\pi \delta\left(\frac{(k_1^2 - k_2^2)}{2m_\chi} + \frac{p_1^2 - (\mathbf{k}_1+\mathbf{p}_1 - \mathbf{k}_2)^2}{2m_{\rm ion}}\right) \ ,
\eea
where in this case
\bea
F(q^0/q) = F_1(A_e) + b_p F_1(A_p) + 4b_{\rm He} F_1(A_{\rm He}) \approx 1 + b_p F_1(A_p) + 4b_{\rm He} F_1(A_{\rm He}) \ .
\eea
Therefore, one finds
\bea
\Gamma_{\chi-{\rm ion}} = \frac{n_i \kappa ^2 e^2 e_D^2 Z_i^2}{(2\pi T)^{3/2} m_i^{1/2}} \int dx_q dc_q \exp\left[ - \frac{1}{2}\left(\frac{1}{2}x_q + A_i\right)^2 \right] \frac{1}{|x_q^2 + \frac{\mu_{\rm Debye}^2}{m_i T} F(q^0/q)|^2} \frac{x_q^5}{(x_q^2 + \frac{m_V^2}{m_i T})^2}, 
\eea
where
\bea
F(q^0/q) = 1 + \sum_i F_1(A_i) \ .
\eea
Both rates, $\Gamma_{\chi-{\rm ion}}$ and $\Gamma_{\chi-e}$, exhibit a strong dependence on scattering angle/momentum transfer.

\subsubsection{Numerical results in the small $m_V$ limit} 

The $m_V \to 0$ case is the most complex since Coulomb scattering is strongly enhanced in the forward region. 
Even with screening, the total scattering rate,
\bea\label{eq:massless2}
\Gamma_{\chi-e} = \frac{n_e \kappa ^2 e^2 e_D^2}{(2\pi T)^{3/2} m_e^{1/2}} \int dx_q dc_q \exp\left[ - \frac{1}{2}\left(\frac{1}{2}x_q + A_e\right)^2 \right] \frac{x_q}{|x_q^2 + \frac{\mu_{\rm Debye}^2}{m_e T} F_1(A_e)|^2} \ ,
\eea
is still too large.
However, since we want the DM particle to be accelerated inside the Sun, we are more interested in large-angle scattering. Therefore, we can add an IR cutoff to $x_q$ that sets
\bea
x_q = \frac{q}{\sqrt{m_e T}}= \frac{\Delta v_1 m_\chi}{\sqrt{m_e T}} > \frac{\zeta v_{10} m_\chi}{\sqrt{m_e T}} = \zeta \frac{m_\chi}{m_e} v_{10} \sqrt{\frac{m_e}{T}} \ ,
\eea
where, in the simulation, the value of $v_{10}$ is fixed to be $7\times10^{-4}~c = 210~{\rm km~ sec^{-1}}$. One observes that the physical meaning of the cut-off is that for each scattering event, we neglect the case that the magnitude of the change of ${\bf v}_1$ is smaller than $\zeta v_{10}$. 

Then, at each step of the simulation, just as in the case of contact interactions, we use the total rate (with the IR cut-off in place) to calculate the scattering probability. If scattering occurs, just as before, we use the differential rate $d\Gamma_{\chi-e}/dc_q dx_q$ to generate the direction and the final momentum of the outgoing dark matter.  

\begin{figure}
\centering
\includegraphics[height=3in]{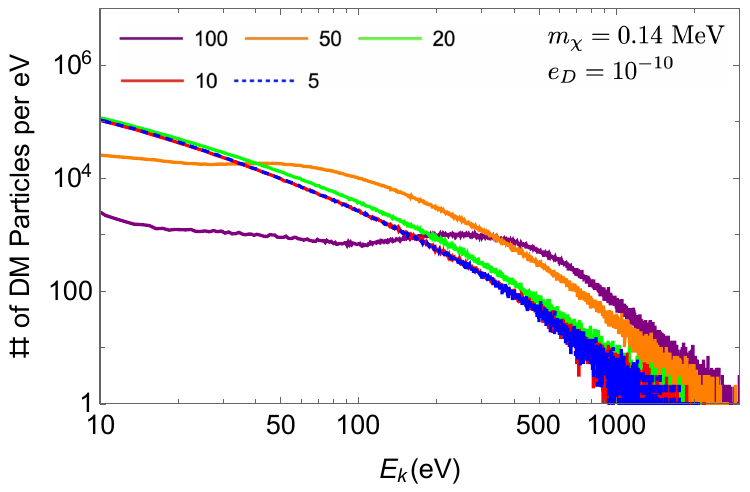}
\caption{The dependence of simulated reflected spectrum on the IR cutoff (labeled by values of $\zeta$). Small values for the cutoff lead to a ``migration" of the hard tail of the reflected spectrum to softer values due to the effective viscosity. The similarity of the blue and red curves indicates that
values of $\zeta < 10$ are adequate for obtaining realistic results. }\label{fig:massless1}
\end{figure}

With the IR cutoff in place, scattering events can be divided into two categories, namely, hard scattering with $x_q > x_q^{\rm cut}$, and soft scattering with $x_q < x_q^{\rm cut}$. Hard scattering events can accelerate the DM particles and generate a hard reflected spectrum, whereas soft scatterings will provide effective friction and soften the reflected spectrum. The smaller the value of $x_q$, the more viscous effects are included in the simulation. In Fig.~\ref{fig:massless1}, we show the reflected spectrum with $m_V = 0$, and $e_D \kappa = 10^{-10}$ for different choices of $\zeta$, from $\zeta = 5$ to 100. The total number of DM particles we shoot into Sun is $10^7$, and the initial conditions are chosen as described in Sec.~\ref{sec:strategy}. For $\zeta = 100$ (as shown by the purple curve), only very hard scattering events remain, and the results are qualitatively similar to the contact interaction case as discussed in the previous sections. One observes clearly that decreasing $x_q$ softens the spectrum and also enhances the soft spectrum overall. Furthermore, we can see that the curves for $\zeta = 5$ and $\zeta = 10$ completely coincide with each other. Therefore, we can conclude that a $x_q$ cut with $\zeta < 10$ is sufficiently accurate for us to capture the relevant soft scattering effects. In practice, our computational power allows us to use $\zeta$ as small as 0.5, which should be able to capture all the soft scattering effects. Comparing the blue dotted curve with the reflected curves in contact scattering case, we observe that in the light mediator case, the spectrum is much softer, as expected. This suggests that the use of detectors with low thresholds would be advantageous in the search for the reflected flux in light mediator DM models.

\subsection{Scattering from the Solar Corona}

While the main focus of our work is a derivation of the sensitivity to the weakest couplings, it is also important to appreciate the limitations of our method when the couplings and the cross sections become large, and the mean free path for $\chi$ becomes so small that DM scatters too quickly to pass into the interior of the Sun. In such a situation, one can entertain the possibility of DM reflecting from the solar corona, that has a small column density, but rather large temperatures (that are indeed $\sim$ two order of magnitude larger than the temperature of the photosphere). The temperature and electron distributions in the solar atmosphere are shown in the lower panel of Fig.~\ref{fig:solarInfo}). We have also simulated the reflected flux from the corona, finding it to be very small, but more energetic than the flux reflected from the photosphere. 

We will consider the possibility of a signal in the SENSEI surface run, which will be discussed in detail in Sec.~\ref{sec:sensei}. In this case, the kinetic energy of the reflected DM needs to be above several~eV. To obtain a qualitative estimate of the reflected spectrum, we can use the approximation described in Sec.~\ref{sec:onescatter}. Then, similar to Eq.~(\ref{eq:311}), one  obtains the differential scattering rate
\bea\label{eq:rateC}
\frac{d\Gamma_{\chi-e}}{d E_2} \approx \frac{\kappa^2 e^2 e_D^2 n_e m_e}{4\pi (2\pi m_e T)^{1/2} m_\chi E_2^2} \exp\left[ - \frac{E_2}{T} \frac{(m_\chi + m_e)^2}{4m_\chi m_e} \right] \ .
\eea 
The Sun's corona extends to radii  $10^6 - 10^7$~km with an electron density smaller than $10^9$ cm$^{-3}$~\cite{2008GeofI..47..197D}. For example, from Eq.~(\ref{eq:rateC}) one concludes that for $\kappa e_D \sim 10^{-4}$, the probability for DM to be reflected with, say, $E_2 > 15$~eV (leading to a multi-electron signal in a silicon detector) within the corona is smaller than one percent. Therefore, we may use the single scattering approximation to estimate the reflected spectrum with the differential scattering probability given by Eq.~(\ref{eq:310}).

\begin{figure}
\centering
\includegraphics[height=3in]{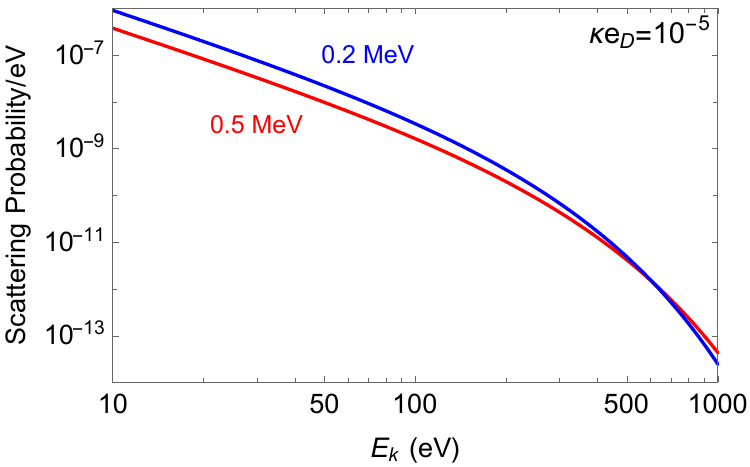}
\caption{Differential scattering probability for $m_\chi = 0.5$ MeV (red) and 0.2 MeV (blue) with $\kappa e_D = 10^{-5}$.}
\label{fig:coronaFlux}
\end{figure}

Due to the Sun's gravity, the trajectories of DM particles are hyperbolic. If the initial impact parameter is small enough the DM particle may hit the surface of the Sun. Once this happens, for a milli-charge $\kappa e_D$ as large as $10^{-5}$ the DM particles can quickly achieve thermal equilibrium with the electrons at the surface of the Sun, and then escape with eV-scale kinetic energy in a random direction. The corona-accelerated DM particles may also hit the surface of the Sun and lose their acquired kinetic energy. Such circumstances complicate the simulation of the accelerated spectrum. As we are primarily interested in a first order estimation of the corona-reflected flux, we will neglect these complications. Concretely, we simply approximate the trajectories as straight lines, and neglect the change of the velocity due to gravity. The differential scattering probability then becomes,
\bea
\frac{dP}{d E_2} \approx \left\langle \frac{1}{v_\chi} \right\rangle \int_0^{R_{\rm c}} \frac{4\pi r^2 dr}{\pi R_{\rm c}^2} \frac{d\Gamma_{\chi-e}}{d E_2} (r) \ ,
\eea
where $R_c$ is the radius enclosed by the corona; we choose $R_c = 1.69\times 10^6$ km (the solar radius is $6.9\times 10^{5}$ km).
The differential scattering probability for $m_\chi = 0.2$ and 0.5 MeV with $\kappa e_D = 10^{-5}$ is shown in Fig.~\ref{fig:coronaFlux}.
As shown in the lower panel of Fig.~\ref{fig:solarInfo}, the coronal temperature remains around $10^6$ kelvin at a height of $10^6$ km. The the index for the power law decay of the density $n_e$ at this height is slower than $-3$. Therefore, we do not expect the contributions to $dP/dE_2$ from the region beyond $r = 1.69\times 10^6$~km to be entirely negligible. 
However, 
given limited data on the coronal temperature and density profiles at even larger radii, 
we limit $R_c = 1.69 \times 10^6$~km as our largest radius. 
In this sense, our sensitivity result is conservative. 

\section{Direct detection of the solar reflected flux}

In this section we shall use the solar reflected DM fluxes and spectrum obtained above to set constraints from direct detection experiments. We focus on the XENON1T experiment, and begin by introducing the standard components that go into the computation of these limits. 

The DM-electron cross section is conventionally written in terms of a reference value, where the DM-electron three-momentum transfer is evaluated at the typical atomic scale $q_0=\alpha m_e$, 
$  \bar \sigma_e \equiv \mu_e^2 
\overline{ |\mathcal{M}(q= q_0)|^2}  / (16\pi m_{\chi}^2 m_e^2 )  $,
where $ \overline{ |\mathcal{M}(q)|^2}$ is the squared matrix element for the
scattering on a free electron and $\mu_e$ is the DM-electron reduced mass. When the interactions are mediated by a kinetically mixed vector (dark-photon), the cross section reads
\begin{align}
  \bar \sigma_e = \frac{16\pi \kappa^2 \alpha \alpha_D \mu_e^2}{(q_0^2 + m_V^2)^2} ,
\end{align}
where $m_V$ is the dark photon mass and $\alpha_D = e_D^2/4\pi$ with $e_D$ the gauge coupling in the dark sector.
The actual dependence of the cross section on momentum transfer is shifted into a ``DM form factor",
\begin{align}
  |F_{\rm DM}(q)|^2\equiv
  \frac{ \overline{ |\mathcal{M}(q)|^2} } { \overline{ |\mathcal{M}(q= q_0)|^2}} =
  \begin{cases}
    1 & \text{contact} \\
    \frac{m_V^2 + q_0^2}{m_V^2 + q^2} & \text{light med.}  \end{cases}\,.
\end{align}
Notably, in the limit $m_V\to 0$ one recovers the millicharged DM model where $Q_{\rm eff} =e_D \kappa/e $ is the fractional charge 
and  $F_{\rm DM}=(\alpha m_e/q)^2$.

\subsection{Limits from XENON1T}

\begin{figure}
\centering
\includegraphics[width=0.5\textwidth]{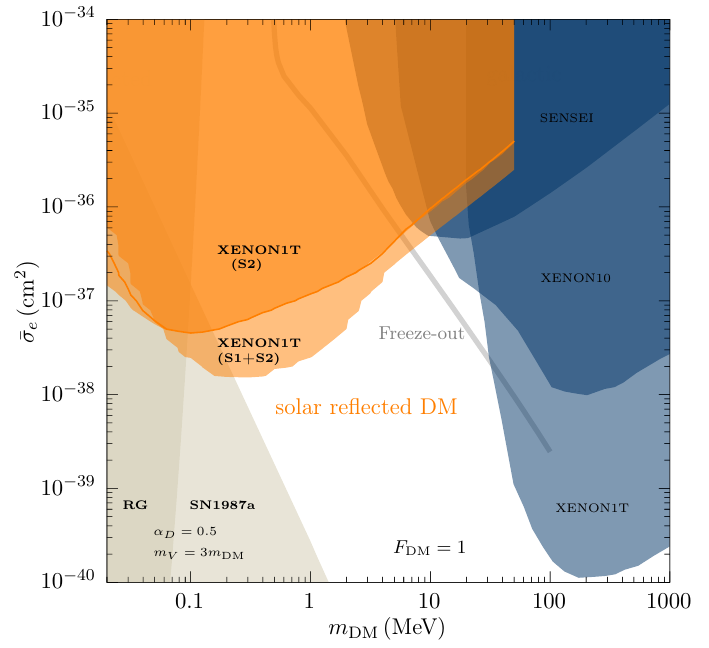}%
\includegraphics[width=0.5\textwidth]{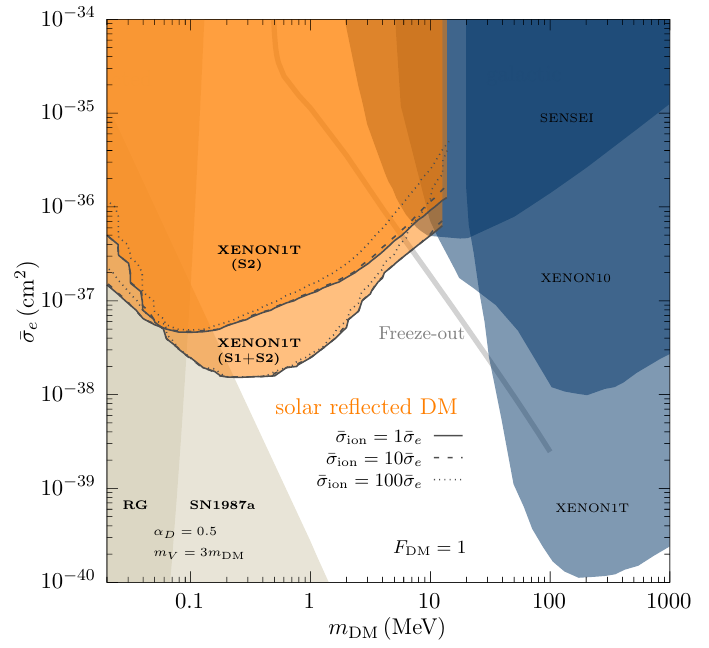}%
\caption{Direct detection limits derived in this work from XENON1T S2~\cite{XENON:2019gfn} and S1+S2~\cite{XENON:2020rca} data from solar reflected DM in the limit of point-like interactions $F_{\rm DM} = 1$ (orange shaded regions as labeled.) The blue shaded regions are previous direct detection limits using the galactic flux component from XENON1T~\cite{XENON:2019gfn}, XENON10~\cite{Essig:2017kqs}, and SENSEI~\cite{SENSEI:2020dpa}. The gray shaded regions to the left are stellar cooling constraints from red giant (RG) stars~\cite{Chang:2019xva} and from the proto-neutron star of SN1987a~\cite{Chang:2018rso} for a dark photon-mediated model with $\alpha_D = 0.5$ and $m_V = 3m_{\rm DM}$. The freeze-out curve is for a complex scalar DM model~\cite{An:2017ojc}.
\textit{Left panel:} only scattering with electrons is assumed; the solid orange line shows the data-driven limit on the charge yield where the lowest energy deposition is 0.19~keV~\cite{LUX:2017ojt}. \textit{Right panel:} various values for the scattering cross section on ions, in relation to electrons are assumed as labeled. The additional scattering channel has the net effect of weakening the limits, because particles are reflected at larger, hence colder, solar radii.}
\label{fig:DDcontact}\end{figure}

\begin{figure}
\centering
\includegraphics[width=0.5\textwidth]{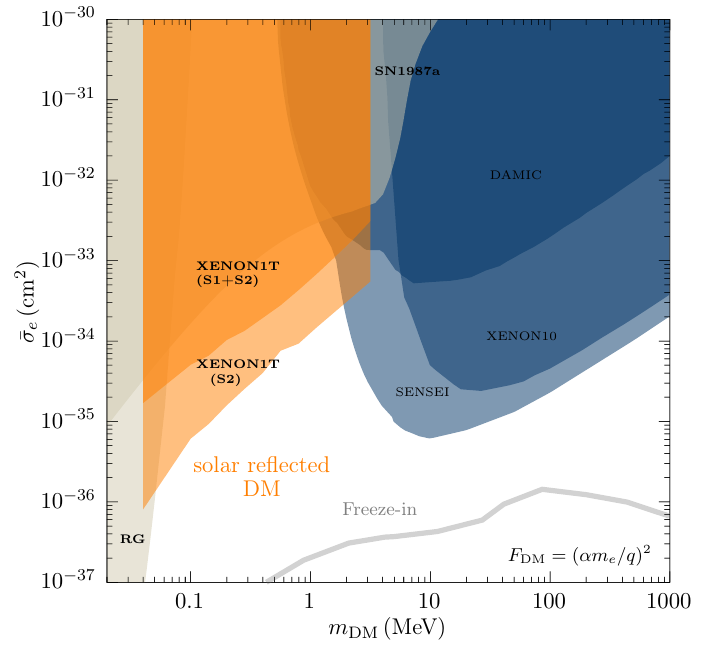}%
\includegraphics[width=0.5\textwidth]{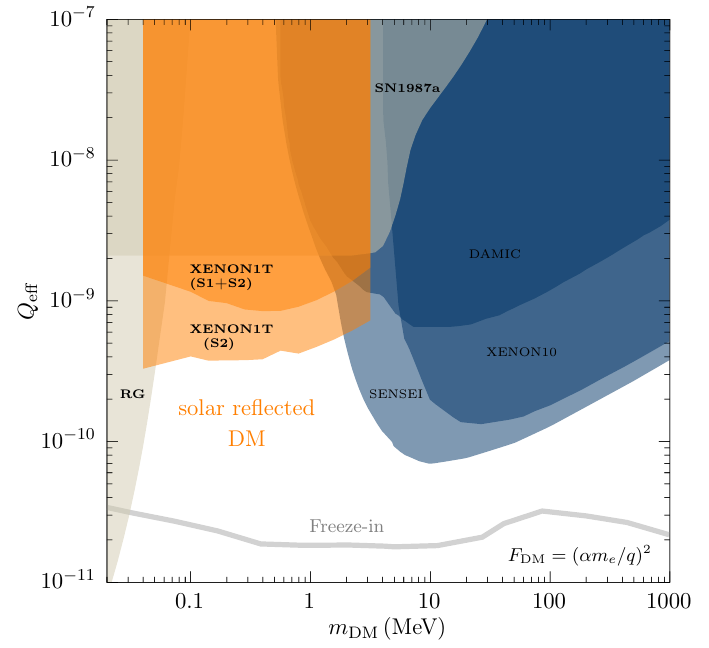}%
\caption{Direct detection limits from solar reflected DM as in Fig.~\ref{fig:DDcontact} but for massless mediator interactions or millicharged DM, $F_{\rm DM} = (\alpha m_e/q)^2$. The standard galactic direct detection limits are from SENSEI~\cite{SENSEI:2020dpa}, XENON10~\cite{Essig:2017kqs}, and DAMIC~\cite{DAMIC:2019dcn}. The astrophysical SN1987a~\cite{Chang:2018rso} and RG~\cite{Vogel:2013raa} cooling constraints are shown as before. The freeze-in line is taken from~\cite{Chang:2019xva}; see also~\cite{Dvorkin:2019zdi}. \textit{Left panel:} constraints in terms of the DM-electron reference cross section evaluated at a momentum transfer $q_0=\alpha m_e$. \textit{Right panel:} constraints in terms of the millicharge measured in units of the elementary charge $Q_{\rm eff}$; in the dark photon model with negligible vector mass, $Q_{\rm eff}= \kappa e_D/e$ holds, where $e_D$ is the gauge coupling in the dark sector.}\label{fig:DDmQ}
\end{figure}

The velocity averaged electron recoil cross section for ionization from an initial atomic state with principal and angular quantum numbers $n$ and $l$ can be written in a standard form~\cite{Essig:2011nj},
\begin{align}
    \label{eq:DDcs}
   \frac{ d \langle\sigma_{n,l} \rangle }{ d\ln E_{e}  }
  & =  \frac{\bar\sigma_e}{8\mu_e^2} \int dq\, \left[ q  |F_{\rm DM}(q)|^2 |f^{\rm ion}_{nl}(p_e, q)|^2 
   \eta(v_{\rm min}(q,\Delta E_{n,l})) \right] .
\end{align}
Here, $|f^{\rm ion}_{nl}(p_e, q)|^2$ is an ionization form factor which depends on momentum transfer $q$ and final state electron momentrum through $E_e = p_e^2/2m_e$. The flux-average of the {\it squared} inverse speed is denoted by $\eta(v_{\rm min})$.%
\footnote{An additional factor $1/v$ is introduced as~\eqref{eq:DDcs} is for $ d \langle\sigma_{n,l} \rangle $ instead of $d \langle\sigma_{n,l} v \rangle $. A factor of $v$ is contained in the overall solar flux that multiplies the ionization cross section in the evaluation of the event rate.}
In the non-relativistic limit it reads,
\begin{align}
\label{eta}
    \eta(E_{ \rm min}) = \int_{E_{\rm min}} dE \,
\frac{m_{\rm DM}}{2E}
    \,\frac{1}{\Phi_{\rm halo} } \frac{d\Phi_{\rm refl}}{dE} \ ,
\end{align}
where $E_{ \rm min}$ is the minimum  incident DM energy required to produce an electron of recoil energy $E_{e}$ after lifting it into the continuum by an investment of the binding energy $|E_{n,l}|$. In~\eqref{eta} we normalize the integral to the halo flux~$\Phi_{\rm halo}$; the total ionization rate from the $(n,l)$ orbital is then given by 
$dR_{nl}/d\ln E_{e} = N_T \Phi_{\rm halo} d \langle \sigma_{nl} \rangle / d\ln E_{e} $, where $N_T$ is the target density.

The energy deposition $\Delta E = E_e + |E_{n,l}|$ induces an ionization signal in xenon~(S2), and, for $E\gtrsim1$~keVee, the scintillation signal~(S1). In this work we use XENON1T S1+S2 data from \cite{Aprile:2020tmw} and the S2-only data from~\cite{Aprile:2019xxb}. We follow our previous work in modeling the formation of the S1 and S2 signals and limit setting procedure~\cite{An:2017ojc}. For the ionization form factor $|f^{\rm ion}_{nl}(p_e, q)|^2$ we use the calculation described in~\cite{Essig:2019xkx}. It is based on numerical solutions of the Schroedinger equation for bound and continuum electrons, and hence yields appropriate results accross the entire momentum-transfer regime. 

The direct detection limits from solar reflection for contact interactions ($F_{\rm DM} = 1$) are shown in Fig.~\ref{fig:DDcontact} with the orange shaded exclusion regions for XENON1T S2~\cite{XENON:2019gfn} and S1+S2~\cite{XENON:2020rca}  as labeled. As can be seen, they attain their maximum strength for $m_{\rm DM}\simeq m_e$ corresponding to the point of most efficient energy transfer in the sun. A variety of complementary constraints are described in the figure caption. 
The limiting constraint (solid orange line) shows the data-driven limit on the charge yield where we demand an energy deposition of at least 0.19~keV congruent with available data~\cite{LUX:2017ojt}.
As shown in the right panel, the additional scattering on ions has the net effect of weakening the limits, because particles are reflected at larger, hence colder, solar radii.

It is instructive to compare the peak sensitivities of XENON1T to galactic dark matter (blue region)
with the sensitivity to reflected dark matter (orange region). 
The strongest constraints on galactic dark matter scattering on electrons, $\sigma \sim 10^{-40}\,{\rm cm}^2$, are derived for $m_\chi \sim 200-300$\,MeV. For the reflected dark matter flux, the best sensitivity is for $\sigma \sim 5\times 10^{-38}\,{\rm cm}^2$ for $S1+S2$ and $\sigma \sim 10^{-37}\,{\rm cm}^2$ for S2 only at masses that are three orders of magnitude smaller, $m_\chi \sim 100-300$\,keV. 
Notice that the lowering of sensitivity by a factor of $10^3$ is composed from the two main factors, large and small.  The geometric loss, proportional to the inverse solid angle subtended by the Sun, $1/\Delta\Omega_\odot\sim 10^6$, is mitigated by the much larger number densities of dark matter at $m_\chi \sim 100-300$\,keV, leading to an overall loss in sensitivity by only a factor of $10^3$. (This rough comparison is possible because the same experiment is used in both cases, XENON1T.)

Results for the millicharged case ($F_{\rm DM} = (\alpha m_e/q)^2$) are shown in Fig.~\ref{fig:DDmQ}. Complementary limits are again described in the figure caption.
In the left panel, constraints in terms of the DM-electron reference cross section is shown. In the right panel, the same constraints are shown in terms of the fractional elementary charge~$Q_{\rm eff} = \kappa e_D/e$. We observe that the limits of $Q_{\rm eff}$ are strong, and relatively mass-independent, surpassing the SN bounds by a slim margin. It is easy to see that the best limits are imposed by the S2 only signal, and this is a direct consequence of the softness of the reflected spectrum in this case. 
Of course, for large enough couplings, the direct detection constraints will eventually disappear as one enters a regime where the reflected flux---prior to entering Earth---will approach a Maxwellian with a temperature resembling the comparatively colder solar surface conditions. 
The precise location of this ceiling is difficult to estimate, but inconsequential as it finds itself in the region that is already excluded by the SN bound ($Q_{\rm eff} \gtrsim 10^{-9}$). In the following section, however, we shall consider reflection from the solar corona as a potential bypass to those conclusions for the large-coupling limit.

\subsection{Detecting the reflected flux from the solar corona}
\label{sec:sensei}

\begin{figure}
\centering
\includegraphics[width=0.5\textwidth]{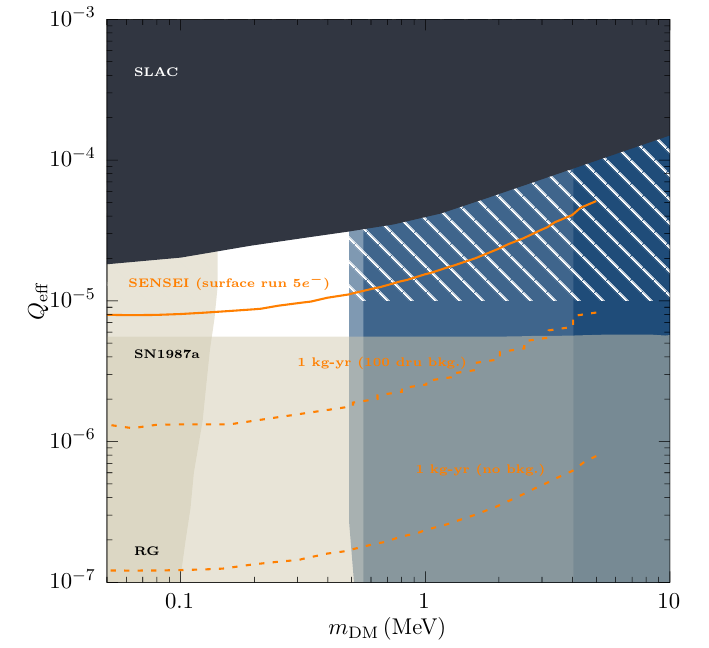}%
\caption{
An extension of the right panel of Fig.~\ref{fig:DDmQ} to larger effective couplings~$Q_{\rm eff}$. The currently allowed white region for $Q_{\rm eff}\gtrsim 5\times 10^{-6}$ opens up from below due to trapping of DM inside the SN, while it is bounded from above by the SLAC millicharge experiment~\cite{Prinz:1998ua}. A surface run by the SENSEI experiment with a small exposure of 0.019~g-days~\cite{Crisler:2018gci} and zero events in the $5e^-$ bin allows us to place a superseded constraint using DM reflected from the solar corona (solid orange line.) A projection to 1~kg-yr exposures assuming a background rate of $100$/kg/day (zero) in the $5e^-$ bin gives the upper (lower) orange dashed curves. The lower curve can be regarded as the most optimal reach for a SENSEI-type surface detector; the actual sensitivity is likely to be limited significantly by backgrounds. The galactic DM constraints obtained with underground searches are the same as in Fig.~\ref{fig:DDmQ}, but disappear for values in excess of $Q_{\rm eff}\sim 10^{-5} $~\cite{Emken:2019tni} as indicated by the hatching.}\label{fig:DDcorona}
\end{figure}

We close the exploration of direct detection sensitivity by commenting on the possibility of probing larger cross sections. In particular, for millicharged DM there exists a currently unconstrained region above the SN1987a bound $Q_{\rm eff}\simeq 5\times 10^{-6}$~\cite{Chang:2018rso}. This is due to the trapping of pair-produced $\chi$-particles inside the supernova. The region has a ceiling at $Q_{\rm eff}\lesssim \text{few}\times 10^{-5}$ due to the direct search at the SLAC millicharge experiment~\cite{Prinz:1998ua}. Both bounds are independent of the DM-nature of~$\chi$. 
In this parameter region, reflected DM experiences re-scattering in the Earth's atmosphere and in any overburden of a direct detection experiment. In particular, we do not expect limits from the XENON1T experiment, as the diffusive process softens and diminishes the spectrum too much. Prospects for probing this region may nevertheless exist through the operation of various low-threshold detectors above ground.

For example, the SENSEI semiconductor direct detection experiment had a surface commissioning run with a very small exposure of $0.019$~gram-days~\cite{Crisler:2018gci}. The threshold for ionizing an electron is $E_{\mathrm{gap}}=1.11~\mathrm{eV}$ and it takes an average of $\varepsilon=3.6~\mathrm{eV}$ for the creation of an electron-hole pair (at least, as measured in high-energy recoils). For example, the corona-reflected flux is capable of creating a charge multiplicity of at least $5e^-$ in a scattering with $E_e \ge 15.5$~eV electron recoil energy, and for which the SENSEI run did not observe any events; (the integer charge multiplicity may be estimated as $Q\left(E_{e}\right)=1+\left\lfloor\left(E_{e}-E_{\mathrm{gap}}\right) / \varepsilon\right\rfloor$~\cite{Essig:2015cda}). 

In Fig.~\ref{fig:DDcorona} we show the resulting limit for the SENSEI surface run using the $5e^-$ bin which had zero counts (orange solid line). As can be seen, it is superseded by the SLAC experiment. However, we also show the results of two projections for an effective exposure of $1$~kg-day for a SENSEI-type experiment. For the upper dashed curve, we assume a background-rate in the $5e^-$ bin of $100$~/kg/day (dru). This appears to be a reasonable ballpark figure based on a simple extrapolation of the $2,3,$ and 4~electron rates for the  SENSEI surface-run. The lower dashed curve assumes (an unrealistic) zero observed events in the $5e^-$ bin. This should therefore be regarded as the optimal reach for a SENSEI-type surface detector. We obtain these lines assuming no re-scattering in the Earth's atmosphere. We caution that this approximation may be subject to corrections, especially for the line of current sensitivity.  The evaluation of semiconductor ionization rates, in a generalization of~\eqref{eq:DDcs} to crystals, follows~\cite{Essig:2015cda} from where we also take the ionization form factor. Complementary constraints from other sources are again described in the figure caption.

\section{Concluding Remarks}

The fast flux of MeV-scale dark matter reflected from the Sun is an inevitable consequence of scattering on solar electrons. This sub-dominant, but highly energetic component of the total dark matter flux on Earth provides a novel pathway to study light dark matter scattering on electrons. While the viability of low mass DM motivates R\&D involving new materials and new ways of registering ever lower recoil energies in direct detection relevant for lower values of $m_\chi$ \cite{Knapen:2017xzo}, the solar reflection mechanism provide a nontrivial level of sensitivity using data from the largest and cleanest dark matter detectors based on Xenon dual phase technology. As a consequence, dark matter electron scattering limits can be imposed {\em now}, without relying on the success of future experimental efforts. A galactic DM particle with mass in the 100 keV to 1 MeV  range has sub-eV kinetic energy and falls below the experimental threshold of any existing experiment. The reflected flux provides a concrete means of probing this part of the parameter space, and current data from XENON1T imposes limits down to cross sections of $10^{-38}$cm$^2$  in the case of contact scattering on electrons. %

While the results for contact DM-electron scattering are an update to our earlier work \cite{An:2017ojc} (see also \cite{Emken:2021lgc}), the derivation of the reflected  flux for DM interactions induced by an ultra-light mediator, and the resulting limits from direct detection presented in this paper, represent a nontrivial extension. The method can now be applied to the most viable sub-MeV DM models, which achieve their sub-thermal relic abundance via freeze-in. The technical challenge to overcome was the strong enhancement of small angle scattering, which leads to the predominant occurrence of soft collisions. The forward singularity is of course regularized by Debye screening. In this paper, we have described a consistent derivation of this screening for the relevant range of energy and momentum transfers, $q^0,q$, and implemented this approach in simulations to generate the reflected spectrum. The numerical treatment was made viable by introducing an artificial IR cutoff, which we varied to confirm that the final spectrum was independent of this parameter. 

The reflected spectrum for a nearly massless mediator is considerably softer than in the case of contact interaction. This is expected, given that frequent soft scatterings downgrade the energetic component of the spectrum. As a result, most of the reflected spectrum leads to ionization in XENON1T that falls below the threshold for registering the S1 signal, and the best constraints are imposed by examining the S2 only part of the signal. The resulting limits in the mass interval for $m_\chi$ from 0.1-to-1 MeV vary from 
$6\times 10^{-36}$cm$^2$ to $10^{-34}$cm$^2$, 
when the cross sections are evaluated at $q=\alpha m_e$. The corresponding limits on the effective charge are around the %
$ Q_{\rm eff} = 4\times  10^{-10}$
benchmark, and are slightly stronger than the conservative limits from analysis of SN cooling. These are the most sensitive constraints by some margin on DM in the sub-MeV mass range obtained using terrestrial probes. Despite the strength of these bounds, we find that the freeze-in prediction for $Q_{\rm eff}(m_\chi)$ is beyond reach using the reflected flux with existing technology.\footnote{Note however, that this need not imply an overproduction of dark matter as the $\chi\bar \chi \to 2V $ annihilation channel remains open, and with the right choice of $e_d$ can ensure correct DM relic abundance for the parameters at the current level of sensitivity.}

\bigskip
{\it Note Added:} As this paper was being finalized, we became aware of recent complementary work exploring cosmological constraints on millicharge DM \cite{Buen-Abad:2021mvc,Nguyen:2021cnb}. The resulting sensitivity to the electron scattering cross section of sub-MeV dark matter is quite similar. However, the results are complementary as the direct detection limits on reflected DM degrade rather weakly in the case of  fractional DM abundance, {\em i.e.} even a sub-1\% fraction of the total DM will have its scattering on electrons constrained via solar reflection, while cosmological bounds are fully relaxed.

\acknowledgments

We thank T.~Emken, R.~Essig and H.~Xu for checking our published results which led to  corrections in the calculation for the reflected flux and DM-electron recoil spectra. The limits have consequently become stronger.
 HA is supported by NSFC under Grant No. 11975134, the National Key Research
and Development Program of China under Grant No. 2017YFA0402204 and the Tsinghua University Initiative Scientific Research Program. MP is supported in part by U.S. Department of Energy (Grant No. DE-SC0011842). JP is supported by the New Frontiers Program of the Austrian Academy of Sciences and by the Austrian Science Fund (FWF) Grant No.~FG~1. AR is supported in part by NSERC, Canada.

\appendix

\section*{Appendices}
\section{Momentum transfer-dependent screening from electrons}
\label{sec:screening}

The Boltzmann equation for the distribution of non-relativistic electrons can be written as
\bea\label{eq:boltzmann}
D f = \frac{\partial f}{\partial t} + \mathbf{v}\cdot \frac{\partial f}{\partial \mathbf{x}} + \dot{\mathbf{k}} \cdot \frac{\partial f}{\partial \mathbf{k}} = 0 \ ,
\eea
where we neglect collisions. The solution takes the form $f = f^{(0)} \propto e^{- k^2/2m_e T}$. We now place the system in a weak and slowly varying electromagnetic field, so that
\bea
\dot{\mathbf{k}} = e(\mathbf{E} + \mathbf{v}\times \mathbf{B}) \ .
\eea
To study the response, $f$ can be expanded as follows 
\bea
f = f^{(0)} + f^{(1)} \ ,
\eea
so the first order expansion of (\ref{eq:boltzmann}) becomes 
\bea
\frac{\partial f^{(1)}}{\partial t} + \mathbf{v}\cdot \frac{\partial f^{(1)}}{\partial \mathbf{x}} = -e (\mathbf{E} + \mathbf{v}\times \mathbf{B}) \cdot \frac{\partial f^{(0)}}{\partial \mathbf{k}} =  -e \mathbf{E} \cdot \hat{\mathbf{k}} \frac{\partial f^{(0)}}{\partial k}   \ ,
\eea
where we use the fact that $f^{(0)}$ is isotropic in $\mathbf{k}$. We will also neglect the contribution from the magnetic field due to the small velocity. Then the solution of $f^{(1)}$ can be formally written as
\bea
f^{(1)}({\bf x, k},t) = - e \frac{\partial f^{(0)}}{\partial k} \hat{\bf k} \cdot \int_{-\infty}^{t} d t' \, {\bf E}(t',{\bf x}- {\bf v}(t - t')) e^{\eta (t'-t)} \ ,
\eea
where $\eta$ is a positive infinitesimal number. The induced electric current is given by
\bea
{\bf j}_{\rm ind}(t,{\bf x}) &=& e \int\frac{d^3k}{(2\pi)^3} {\bf v} f^{(1)}({\bf x},{\bf k},t) \nn
&=& - e^2 \int\frac{d^3k}{(2\pi)^3} \frac{d f^{(0)}(k)}{d k} {\bf v} \int_{-\infty}^{t} d t' \, \hat{\bf k}\cdot {\bf E}(t',{\bf x}- {\bf v}(t - t')) e^{\eta (t'-t)} \nn
&=& - e^2 \int\frac{d^3k}{(2\pi)^3} \frac{d f^{(0)}(k)}{d k} {\bf v} \int_{0}^{\infty} d \tau \, \hat{\bf k}\cdot {\bf E}(t-\tau,{\bf x}- {\bf v}\tau) e^{-\eta \tau} \ .
\eea
We will need the Fourier transform of ${\bf j}_{\rm ind}$ which can be written as follows,
\bea
{\bf j}_{\rm ind}(q^0, {\bf q}) &=& \int dt\, d{\bf x}\, e^{iq^0 t- i {\bf q\cdot x}} {\bf j}_{\rm ind}(t,{\bf x}) \nn
&=& - e^2 \int_0^\infty d\tau \int dt ~e^{iq^0 t- i{\bf q\cdot v} \tau} \int\frac{d{\bf k}}{(2\pi)^3} \frac{d f^{(0)}}{d k} {\bf v}  \left[\hat{\bf k}\cdot {\bf E}(t - \tau,{\bf q})\right] e^{-\eta \tau} \nn
&=& - e^2 \int d\tau e^{(i q^0 - i {\bf q\cdot v} - \eta)\tau} \int\frac{d{\bf k}}{(2\pi)^3} \frac{d f^{(0)}}{d k} {\bf v} \left[ \hat {\bf k}\cdot{\bf E}(q^0, {\bf q}) \right] \nn
&=& \int\frac{d{\bf k}}{(2\pi)^3} \frac{e^2}{ i q^0 - i {\bf q\cdot v} - \eta}  \frac{d f^{(0)}}{d k} \frac{\bf k}{(k^2 + m_e^2)^{1/2}} \left[ \hat {\bf k}\cdot{\bf E}(q^0, {\bf q}) \right] \nn
&\approx& \int\frac{d{\bf k}}{(2\pi)^3} \frac{d f^{(0)}}{d k} \frac{(-i )e^2}{ q^0 - {\bf q}\cdot {\bf k}/m_e + i \eta}   \frac{\bf k}{m_e} \left[ \frac{\bf k}{k}\cdot{\bf E}(q^0, {\bf q}) \right], \nn
\eea
where we use the non-relativistic approximation in the last step. 

The number density of electrons and their phase space distribution function in the zeroth order are given by 
\bea
n_e = \int\frac{d{\bf k}}{(2\pi)^3} f^{(0)}(k) \ ,
~~
f^{(0)}(k) = n_e \left(\frac{2\pi}{m_e T}\right)^{3/2} e^{- k^2/2m_e T} \ ,
\eea
They determine the polarization,
\bea
{\bf P} = \frac{i}{q^0} {\bf j}_{\rm ind} = -e^2 n_e \frac{(2\pi)^{3/2}}{(m_e T)^{5/2}} \int\frac{d{\bf k}}{(2\pi)^3} e^{-k^2/2m_e T} \frac{{\bf k}[{\bf k}\cdot {\bf E}(q^0,{\bf q})]}{m_e q^0(q^0 - {\bf q\cdot k}/m_e + i \eta)} \ .
\eea
Therefore, the permittivity tensor can be written as follows,
\bea
\varepsilon^{ij} = \delta^{ij} - \frac{e^2 n_e}{q^0} \frac{(2\pi)^{3/2}}{m_e^{7/2} T^{5/2}}  \int\frac{d{\bf k}}{(2\pi)^3} e^{-k^2/2m_e T} \frac{k^i k^j}{(q^0 - {\bf q\cdot k}/m_e + i \eta)} \ .
\eea
To determine the Debye screening scale we extract the longitudinal part of $\varepsilon^{ij}$ 
\bea\label{eq:epsL}
\varepsilon_L = \varepsilon^{ij} \frac{q^i q^j}{q^2} = 1 - \frac{e^2 n_e}{T} \frac{1}{m_e^2} \left(\frac{2\pi}{m_e T}\right)^{3/2} \int\frac{e^{-k^2 /m_e T}k^2 dk\, d\cos\theta}{(2\pi)^2} \frac{k^2 \cos^2\theta}{q^0(q^0 - k q\cos\theta/m_e + i\eta) } \ ,\nn
\eea
and perform the integration over $\cos\theta$:
\bea
\int d\cos\theta \frac{\cos^2\theta}{q^0 - kq\cos\theta/m_e + i\eta} = \frac{1}{q^0 + i\eta} \left[  \frac{2}{b^3}{\rm tanh}^{-1}[b] - \frac{2}{b^2} \right] \ ,
\eea
where 
\bea
b = \frac{kq}{m_e( q^0+ i\eta)} \ .
\eea
Therefore, we obtain
\bea
\varepsilon_L = 1 + e^2 n_e \left(\frac{2\pi}{m_e T}\right)^{3/2} \frac{1}{m_e^2 T}\int \frac{k^4 dk}{(2\pi)^2} \frac{1}{q^0 (q^0+ i\eta)} \left[\frac{2}{b^2} - \frac{2}{b^3} \tanh^{-1}b\right] e^{-k^2/2m_e T}\ .
\eea
In the limit of vanishing energy transfer, $q^0 = 0$, we should recover the standard literature result. It turns out, that the correct procedure is to take the $\eta\rightarrow0$ limit first. Then, as $q^0\rightarrow0$, all $q^0$ factors cancel, and we are left with
\bea
\varepsilon_L &\rightarrow& 1 + e^2 n_e \left(\frac{2\pi}{m_e T}\right)^{3/2} \frac{1}{m_e^2 T} \int \frac{k^4 dk}{(2\pi)^2} e^{-k^2/2m_e T} 2\left(\frac{m_e}{k q}\right)^2  \nn
&=& 1 + \frac{e^2 n_e}{T q^2} \ ,
\eea
which agrees with the standard electrostatic formula. 

\begin{figure}
\centering
\includegraphics[height=3in]{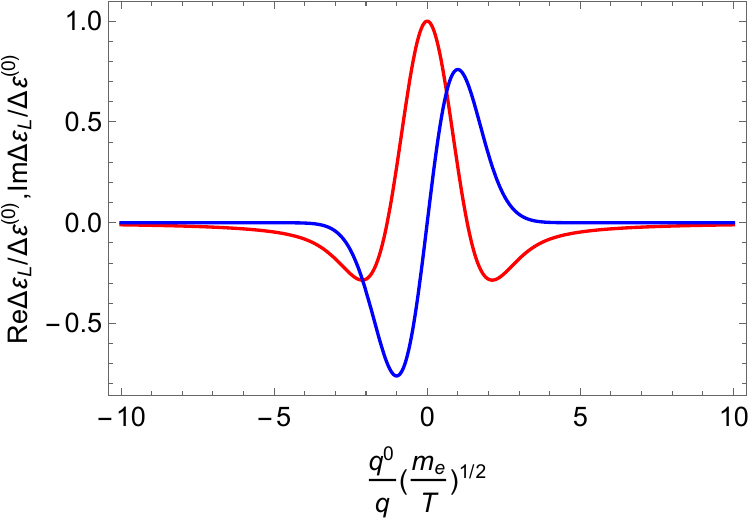}
\caption{Dispersive (red) and absorptive (blue) parts of the longitudinal permittivity $\varepsilon_L$ as a function of the parameter $A$.}\label{fig:debye}
\end{figure}

For $q^0 > 0$, we first neglect all the $\eta$'s outside the logarithm, and obtain 
\bea\label{eq:epsilonL}
\varepsilon_L &=& 1 + \frac{e^2 n_e}{T q^2} \left[1 - 2A \int \frac{x dx}{(2\pi)^{1/2}} \tanh^{-1} (x/A) e^{-x^2/2} \right] \nn
&\equiv& 1 + \frac{e^2 n_e}{T q^2} F_1(A),
\eea
where 
\bea\label{eq:Ae}
A = \frac{q^0}{q} \left(\frac{m_e}{T}\right)^{1/2}
\eea
and 
\bea
F_1(A) = 1 - \left(\frac{\pi}{2}\right)^{1/2} A e^{-A^2/2} \left[{\rm erfi}(A/\sqrt{2}) - i\right] \ .
\eea
The real and imaginary parts of $F_1$ are shown in Fig.~\ref{fig:debye}, which indicate that the correction from non-zero $q^0$ can be substantial and
of order unity.

Similarly, we may derive the correction to the transverse mode. We project $\varepsilon^{ij}$ onto the transverse polarization vector, and exploiting the fact that the plasma is isotropic and homogeneous, take the average of the two polarizations. This way we obtain
\bea
\varepsilon_T = 1 - \frac{e^2 n_e}{T} \frac{1}{m_e^2} \left(\frac{2\pi}{m_e T}\right)^{3/2} \int\frac{e^{-k^2 /m_e T}k^2 dk d\cos\theta}{(2\pi)^2} \frac{\frac{1}{2} k^2 \sin^2\theta}{q^0(q^0 - k q\cos\theta/m_e + i\eta) } \ .\nn
\eea
On computing the $\theta$-integral, we find 
\bea
\varepsilon_T &=& 1 - \frac{e^2 n_e}{T} \frac{1}{m_e^2} \left(\frac{2\pi}{m_e T}\right)^{3/2} \int\frac{e^{-k^2 /m_e T}k^4 dk}{(2\pi)^2 q^0 (q^0 + i\eta)} \frac{(b^2 - 1)\tanh^{-1}[b] + b}{b^3} \nn
&=& 1 - \frac{e^2 n_e}{T q^2} A \int \frac{x dx}{(2\pi)^{1/2}} e^{-x^2/2} \left[ \left(\frac{x^2}{A^2} - 1\right)\tanh^{-1}\left(\frac{x}{A}\right) + \frac{x}{A} \right] \nn
&=& 1 - \frac{e^2 n_e}{T q^2} \frac{1}{A} \left(\frac{\pi}{2}\right)^{1/2} e^{-A^2/2} \left[{\rm erfi}\left(\frac{A}{\sqrt{2}}\right) - i\right] \ .
\eea

In the non-relativistic limit, for small $q$, $\varepsilon_T$ should have a zero at $(q^0)^2 = \omega_p^2$, where $\omega_p^2 = e^2 n_e/m_e$ is the plasma frequency. It is easy to see that this is indeed the case by taking the limit $A\gg 1$:
\bea
\left(\frac{\pi}{2}\right)^{1/2} \frac{1}{A} e^{-A^2/2} {\rm erfi}(A/\sqrt{2})  \rightarrow \frac{1}{A^2} \ .
\eea
Therefore, one finds the expected pole in this limit
\bea
\varepsilon_T \rightarrow 1 - \frac{e^2 n_e}{m_e {q^0}^2} \ .
\eea
Similarly, one can find the large $A$ limit of $\varepsilon_L$, which coincides with the expression above. 

\begin{figure}
\centering
\includegraphics[height=3in]{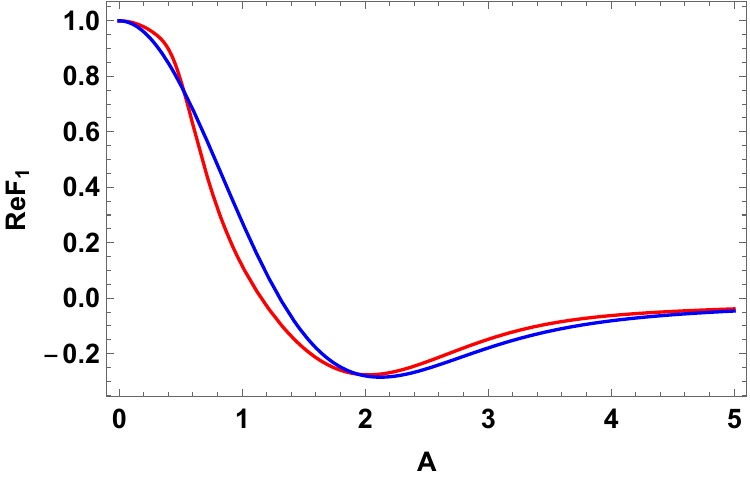}
\caption{Comparison between ${\rm Re} F_1$ and ${\rm Re} F_1'$.}\label{fig:ReF1p}
\end{figure}

Finally, we would like to comment on a numerical subtlety in implementing $F_1(A)$ in simulations. 
The real part of $F_1$ contains the error function with an imaginary argument, which is not defined in ROOT. Therefore, in order to have a feasible numerical scheme, we use the following empirical formula, 
\bea
{\rm Re}F_1' = \left(1- \frac{x^3}{5}\right)\exp\left(-\frac{x^2}{2}\right)- \frac{1}{x^2}\exp\left(-\frac{1}{x^2}\right) \ .
\eea
A comparison between ${\rm Re}F_1$ and ${\rm Re} F_1'$ indicating their similarity is shown in Fig.~\ref{fig:ReF1p}. 

%\bibliography{refs}

\end{document}